\documentclass[prb,twocolumn,amsmath,amssymb,floatfix,showpacs,superscriptaddress]{revtex4}

\usepackage{bm}
\usepackage[usenames,dvips]{color}
\usepackage{epsfig}

\newcommand{\up}{\uparrow}
\newcommand{\down}{\downarrow}
\def\nab{{\mbox{\boldmath{$\nabla$}}}}
\def\sig{{\mbox{\boldmath{$\sigma$}}}}

\newcommand{\equref}[1]{Eq.~\eqref{#1}}
\newcommand{\appendref}[1]{Appendix~\ref{#1}}
\newcommand{\figref}[1]{Fig.~\ref{#1}}
\newcommand{\secref}[1]{Sec.~\ref{#1}}
\newcommand{\tableref}[1]{Table~\ref{#1}}

\begin{document}
\title{Pair-breaking effect on mesoscopic
persistent currents}

\author{H. Bary-Soroker}
\email{hamutal.soroker@weizmann.ac.il}

\affiliation{Department of Condensed Matter Physics, Weizmann
Institute of Science, Rehovot 76100, Israel}

\author{O. Entin-Wohlman}

\affiliation{Department of Physics, Ben Gurion University, Beer
Sheva 84105, Israel}

\affiliation{Albert Einstein Minerva Center for Theoretical
Physics, Weizmann Institute of Science, Rehovot 76100, Israel}

\author{Y. Imry}

\affiliation{Department of Condensed Matter Physics, Weizmann
Institute of Science, Rehovot 76100, Israel}

\date{\today}

\begin{abstract}
We consider the contribution of superconducting
fluctuations to the mesoscopic persistent current (PC) of an
ensemble of normal metallic rings, made of a superconducting
material whose low bare transition temperature $T^{0}_{c}$ is
much smaller than the Thouless energy $E_{c}$. The effect of
pair breaking is introduced via the example of magnetic
impurities. We find that over a rather broad range of
pair-breaking strength $\hbar/\tau_{s}$, such that $T_c^0
\lesssim \hbar/\tau_s \lesssim E_c$, the superconducting
transition temperature is normalized down to minute values or
zero while the PC is hardly affected. This may provide an
explanation for the magnitude of the average PC's in copper and
gold, as well as a way to determine their $T^0_c$'s. The
dependence of the current and the dominant superconducting
fluctuations on $E_c\tau_s$ and on the ratio between $E_c$ and
the temperature is analyzed. The measured PC's in copper (gold)
correspond to $T^0_c$ of a few (a fraction of) mK.
\end{abstract}

 \maketitle

\section{Introduction}\label{section:Introduction}

Equilibrium persistent currents (PC's), flowing in normal
mesoscopic metallic rings, have been a challenge for both
experimentalists and theorists. The persistent current is a
manifestation of the Aharonov-Bohm effect: it appears when the
ring is threaded by a magnetic flux, and it is periodic in the
flux enclosed in the ring. \cite{BIL,book} Due to
energy-averaging and phase-coherence limitations, one expects to
monitor in experiment only the lowest harmonics in the flux
quantum $h/e$.

Surprisingly enough, the magnitudes of the PC's measured on huge
collections of rings ($10^7$ copper rings \cite{LDDB} and $10^5$
silver rings \cite{DBRBM}) turned out to be larger than those
expected theoretically. The periodicity observed in these large
ensembles is $h/2e$, i.e., half of the magnetic flux quantum. On
the other hand, measurements on a single
ring\cite{Bluhm_Koshnick_Bert_Huber_Moler,Chandrasekhar_Webb_et_al}
or on a small number\cite{JMKW} of gold rings showed the $h/e$
periodicity. In the collection of 30 gold rings\cite{JMKW} both
the the $h/2e$ harmonic and the $h/e$ one were observed.
Overall, the sign of the amplitude of the $h/2e$ harmonic
measured on metallic rings seems to indicate that the low-flux
response is diamagnetic.\cite{DBRBM,JMKW}

In the experiments on ensembles of rings,\cite{LDDB,DBRBM,JMKW}
the average PC was found by measuring the magnetic moment
produced by all rings, which was then divided by the number of
rings, $N$, to yield the net average current of a single ring.
In most of the
experiments\cite{LDDB,DBRBM,Bluhm_Koshnick_Bert_Huber_Moler,JMKW}
the magnitude of the average PC, at low temperatures, is roughly
of the order of $eE_c/\hbar$. Here $E_c = \hbar D/L^2$ is the
Thouless energy, $L$ is the circumference of the ring and $D=v_F
l_{\textrm{el}}/3$ is the diffusion coefficient, where
$l_{\textrm{el}}$ is the elastic mean free path, and $v_F$ is
the Fermi velocity. (We consider the diffusive, $L\gg
l_{\textrm{el}}$, case.)

The first theoretical studies of the PC have been carried out on
grand-canonical systems of non-interacting electrons.\cite{book,CGR}
In these theories, the current in each ring is $h/e$ periodic. The
sign and magnitude of the PC of the individual rings vary randomly
due to their high sensitivity to the disorder and to the system's
size. This results in a very small average PC, which is dominated by
the exponential factor $\exp(-L/2l_{\textrm{el}})$. Hence, the
typical magnitude of the current is predicted to be $\sqrt{N}$ times
the standard deviation of the PC of non-interacting electrons, which
at low temperatures is of the order of $eE_c/\hbar$. Consequently,
the persistent current carried by non-interacting electrons is too
small to explain the large-ensemble experiments. Similarly, the PC
predicted for non-interacting electrons in the canonical
ensemble\cite{AGI} is substantially too small to explain the
observed amplitude of the $h/2e$ harmonic.

The theory for interacting electron systems\cite{AEPRL,AEEPL}
predicts $h/2e$ periodicity of the interaction-dependent part of the
PC. According to this theory, the average magnitude of the PC per
ring due to interactions is independent of the number of rings. The
total measured PC, divided by $N$, is thus expected to have an
$N$-independent contribution due to interactions, and an
interaction-independent contribution proportional to $N^{-1/2}$. The
presence of the $h/e$ harmonic in the measurements performed on a
single ring
\cite{Bluhm_Koshnick_Bert_Huber_Moler,Chandrasekhar_Webb_et_al} and
on a few\cite{JMKW} rings, and its absence in large
ensembles,\cite{LDDB,DBRBM} are in agreement with these theoretical
predictions. Experiments on a single ring
\cite{Mailly_Chapelier_Benoit} and on a large
ensemble\cite{Reulet_Ramin_Bouchiat_Mailly_and_Deblock_Naot_Bouchiat_Reulet_Mailly}
of semiconducting rings show the $h/e$ and the $h/2e$ periodicities
respectively, consistent with the arguments given above.

Notwithstanding the order of the harmonics, their amplitudes, in
particular that of the $h/2e$ one, remained unexplained for the
large-ensemble measurements. On the other hand, the magnitudes of
the $h/e$ harmonic measured in Refs.
\onlinecite{Bluhm_Koshnick_Bert_Huber_Moler} and \onlinecite{JMKW}
agree roughly with the prediction for non-interacting electrons,
while the PC measured by Chandrasekhar {\it et al.}
\cite{Chandrasekhar_Webb_et_al} turns out to be much larger, however.

Here we study the PC of large ensembles, focusing on the role of
electronic interactions. These, attractive and repulsive of
reasonable strengths, give rise to comparable magnitudes of the
averaged PC (within an order of magnitude), but predict opposite
signs. Whereas repulsive electron-electron interactions \cite{AEPRL}
result in a paramagnetic response at small magnetic fluxes,
attractive interactions yield a diamagnetic response, \cite{AEEPL}
as indeed seems to be indicated in the experiments. The magnitude of
the PC predicted for electrons which interact repulsively is
smaller\cite{COM1} by a factor of about five than, e.g., the
magnitude of the PC measured in copper.\cite{LDDB} The effective
coupling strength of repulsive interactions decreases as the
temperature decreases, due to interactions mediated by states whose
energies are large compared with the temperature.\cite{dG,MA} This
``downwards" renormalization is the reason for the disagreement
between the theory for electrons interacting repulsively and the
experiments.\cite{AEPRL} On the other hand, the attractive
interaction is normalized ``upwards" at low temperatures, \cite{MA}
and eventually leads to a superconducting state. One expects the
magnitude of the averaged PC due to attractive interactions, i.e.,
due to superconducting fluctuations, \cite{AL} to increase with the
strength of the interaction, or alternatively, to decrease as the
(superconducting) transition temperature is reduced. Since the
transition temperatures of metals such as copper, gold, and silver
-- on which the PC has been measured -- are expected\cite{Mota} to be
extremely small or zero, Ambegaokar and Eckern \cite{AEEPL} have
employed in their estimates  small values of the attractive
coupling. Consequently, they came up with a magnitude for the PC
which is again smaller by a factor of order five than the measured
one.\cite{LDDB}

In order to reconcile the relatively large interaction required to
fit the experiments with the apparent absence of a superconducting
transition, we propose that the rings (of e.g., copper) contain a
tiny amount of magnetic impurities. We show that a small
concentration of these pair-breakers may suffice to hinder the
appearance of superconductivity, while hardly affecting the
magnitude of the PC. Indeed, it seems that a small amount of
magnetic impurities is almost unavoidable in metals like copper.
This is suggested by recent experiments, \cite{PGAPEB} aimed to
measure the temperature dependence of the dephasing time in noble
metal samples. Theoretically, one expects \cite{AAK,book} this rate
to vanish as the temperature goes to zero. However, it was found
that the dephasing time may cease to increase below a certain
temperature. This finding was attributed \cite{PGAPEB} to the
presence of a small concentration of magnetic impurities, which was
reported to exist in these samples.

As is well-known, magnetic impurities act as pair breakers,
leading to the vanishing of the transition temperature $T_c$
once the spin-scattering rate $1/\tau_s$ is larger than the {\em
bare} transition temperature of the material {\em without} the
magnetic impurities, $T_c^0$. \cite{AG} At the same time,
superconducting fluctuations can result in a significant PC
provided that the lifetime of a Cooper-pair ($\sim\tau_s$ at low
temperatures) is larger than the time it takes it to encircle
the ring, $\sim\hbar/E_c$. (In the
experiments\cite{LDDB,DBRBM,JMKW} $E_c\sim 10$~mK.) Therefore,
the observation that the PC is almost unaffected by magnetic
impurities while $T_c$ vanishes holds in the range
\begin{align}
T_c^0 \lesssim 1/\tau_s \lesssim E_c \ , \label{1}
\end{align}
(from now on we use units in which $\hbar=1$).

It is instructive to write the above condition in terms of
lengths, for which \equref{1} reads
\begin{align}
L \lesssim L_s \lesssim \xi(0)\ , \label{2}
\end{align}
where
\begin{align}\label{introduction:LsXi0}
L_s=(D\tau_s)^{1/2},\ \textrm{and} \quad \xi(0)=(D/T_c^0)^{1/2}\
.
\end{align}
Here the magnetic-impurities scattering length $L_s$ is the distance
a diffusing electron covers during the time interval $\tau_s$. The
bulk superconducting coherence length, in the absence of magnetic
impurities $\xi(0)$, is the characteristic distance between two
electrons forming a Cooper-pair. At low temperatures a Cooper-pair
fluctuation can propagate a distance of the order of $L_s$ until it
is destroyed due to the scattering by magnetic impurities. When
$L\lesssim L_s$ the pairs are sensitive to the Aharonov-Bohm flux
and consequently contribute significantly to the PC. When pair
breaking occurs on scales smaller than the characteristics distance
between two paired electrons, i.e., when $\xi(0)>L_s$, then the bulk
material would not become a superconductor. Therefore, rings made of
alloys which are not superconducting in the bulk due to pair
breakers, will have PC's due to Cooper-pair fluctuations provided
that \equref{2} is satisfied. We show that the measured amplitude of
the $h/2e$ harmonic in copper\cite{LDDB} and gold\cite{JMKW} rings
can be understood theoretically, assuming a minute, less than one
part per million, concentration of pair breakers. Similar amounts of
magnetic impurities were obtained for the most purified copper and
gold samples in Ref.~\onlinecite{PGAPEB}. We point out that
according to our considerations, the measurement of the PC provides
a way to estimate $T_c^0$, which may well be unreachable by direct
experiments.

This paper is organized as follows. In
\secref{section:Derivation} and \appendref{appendix} we derive
the expression for the PC due to superconducting fluctuations,
taking into account the effect of pair breakers. In
\secref{section:DominantFluctuations} we characterize the
dominant Matsubara frequencies and wave numbers that contribute
to the PC, and discuss the significant harmonics. In
\secref{section:limits} we expand the expression for the PC in
the limits of high and low temperatures. The effect of pair
breaking on the renormalization of the attractive interaction is
discussed in \secref{section:renormalization}. In
\secref{section:compare_experiments} we present a detailed
comparison of our results with the experimental data, and
estimate $T_c^0$ for copper and gold. Finally, the results are
summarized in \secref{section:results}.

In our analysis the effect of pair-breaking is brought about by
the presence of magnetic impurities, disregarding the Kondo
screening of the spins. Obviously one may consider other pair
breakers, such as two-level systems, \cite{IFS} inelastic
scattering,\cite{LeeRead} or magnetic fields.\cite{SO} Other
effects of magnetic impurities have previously been considered in
Ref.~\onlinecite{ES}.

It was suggested by Kravtsov and Altshuler\cite{KA} that the
measured currents have a different source than the equilibrium PC
discussed so far. A non-equilibrium noise, for example, a stray
ac electric field, can cause a dc current by a rectification
effect. In Ref.~\onlinecite{KA} it was shown that the measured
signal\cite{LDDB} may be explained provided that there exists
such a non-equilibrium noise. This mechanism is different than
the one suggested by us.

\section{Derivation of the persistent current}
\label{section:Derivation}

The PC is obtained by differentiating the free energy of electrons
residing in a ring with respect to the magnetic flux enclosed in
that ring. In this section we derive the term in the free energy
which results from superconducting fluctuations. The system consists
of diffusing electrons which interact with each other attractively,
and are also scattered by magnetic impurities that couple to their
spin degrees of freedom. We use the Hamiltonian \cite{AG}
\begin{align}
H&=\int d{\bf r}\Bigl (\psi^{\dagger}_{\alpha}({\bf r}) \Bigl
[({\cal H}_{0}+u_{1}({\bf r}))\delta_{\alpha\gamma}+u_{2}
({\bf r}){\bf S}\cdot\sig^{\alpha\gamma}\Bigr ]\psi^{}_{\gamma}({\bf r})\nonumber\\
&\ \ \ \ \ \ \ \ \ \ \ \ \ -\frac{g}{2}\psi^{\dagger}_{\alpha}
({\bf r})\psi^{\dagger}_{\gamma}({\bf r})\psi^{}_{\gamma}({\bf
r}) \psi^{}_{\alpha}({\bf r})\Bigr )\ ,\label{HAM}
\end{align}
in which the last term represents the attractive interaction,
with coupling $g~(> 0)$. The spin components are $\alpha$ and
$\gamma$, and $\sig$ is the vector of the Pauli matrices. The
free, spin-independent, part of the Hamiltonian is
\begin{align}
{\cal H}_{0}=(-i\nab - [2\pi/L]\phi{\bf {\hat x}})^{2}/2m -\mu \
,
\end{align}
where $m$ is the electron mass, $\mu $ is the chemical
potential, and $\phi$ is the magnetic flux through the ring, in
units of $h/e$. The unit vector ${\bf {\hat x}}$ points along
the circumference of the ring in the anti-clockwise direction.
The scattering, by both nonmagnetic and magnetic ions, is
assumed to result from $N_{i}$ point-like randomly-located
impurities, such that
\begin{align}
&u_{1}({\bf r})+u_{2}({\bf r}){\bf S}\cdot\sig\nonumber\\
&\equiv \sum_{i=1}^{N_{i}}\Bigl (\delta ({\bf r}-{\bf
R}_{i})-\frac{1}{V}\Bigr ) (u_{1}+u_{2}{\bf S}^{}_{{\bf
R}_{i}}\cdot\sig )\ ,
\end{align}
where $V$ is the system volume.

We calculate the partition function ${\cal Z}$ using the method
of Feynman path integrals combined with the Grassmann algebra of
many-body fermionic coherent states, \cite{AS} in which the
superconducting order-parameter is introduced by the
Hubbard-Stratonovich
transformation.\cite{BarySoroker_EntinWohlman_Imry} Details of
this procedure are given in \appendref{appendix}. As is shown
there, the partition function is (the temperature is denoted by
$T$)
\begin{align}
{\cal Z}={\cal Z}_0\prod_{{\bf q},\nu}\Bigl (
1-\frac{gT}{V}\;\Pi({\bf q},\nu) \Bigr )^{-1} \ ,\label{ZwithPi}
\end{align}
where the polarization,\cite{AA}
\begin{align}
&\Pi({\bf q},\nu)= \frac{1}{2} \sum_{\omega}
\varepsilon_{\alpha\gamma} K_{\omega\alpha\gamma}({\bf q},\nu)\ ,
\end{align}
consists of the Cooperon-dominated contributions
\begin{align}
K_{\omega\alpha\gamma}({\bf q},\nu)&= \sum_{{\bf k}_1, {\bf
k}_2}\langle G_{\alpha\alpha '}({\bf k}_1+{\bf q},{\bf k}_2+{\bf
q},\omega+\nu)\nonumber\\
&\times \varepsilon_{\alpha '\gamma '} G_{\gamma\gamma '}(-{\bf
k}_1,-{\bf k}_2,-\omega)\rangle \ .\label{FUNK}
\end{align}
Here $\varepsilon$ is the anti-symmetric tensor,
$\varepsilon_{\alpha\alpha}=0$, and
$\varepsilon_{\up\down}=-\varepsilon_{\down \up}=1$, and $G$
denotes the particle Green function.

In Ref.~\onlinecite{AG} the polarization $\Pi({\bf q}=0,\nu=0)$ was
calculated from the Dyson equation for the Cooperon. Their
calculation can be extended to general ${\bf q},\nu$
\begin{widetext}
\begin{align}
& K_{\omega\alpha\gamma}({\bf q},\nu)\nonumber\\ &= \sum_{\bf k}
\bar G_{\alpha\alpha}({\bf k}+{\bf q},\omega+\nu) \bar
G_{\gamma\gamma}(-{\bf k},-\omega)\big[\epsilon_{\alpha\gamma}
 + N_i \overline{( u_1\delta_{\alpha\alpha'}+u_2{\bf
S}\cdot\sig^{\alpha\alpha'} ) ( u_1\delta_{\gamma\gamma'}+u_2{\bf
S}\cdot\sig^{\gamma\gamma'} )} K_{\omega\alpha'\gamma'}({\bf
q},\nu)\big]\ .\label{K2}
\end{align}
\end{widetext}
Here, $\bar G_{\alpha\gamma}=\delta_{\alpha\gamma}\bar
{G}_{\alpha \alpha}$ is the Green function averaged over the
impurity disorder and spin components (which makes it diagonal in
spin-space). Averaging over the impurity spins,
\begin{align}
&\overline{( u_1\delta_{\alpha\alpha'}+u_2{\bf
S}\cdot\sig^{\alpha\alpha'} ) (
u_1\;\delta_{\gamma\gamma'}+u_2{\bf S}\cdot\sig^{\gamma\gamma'}
)}\nonumber\\&=u_1^2\delta_{\alpha\alpha'}\delta_{\gamma\gamma'}
+\frac{1}{3}\;
S(S+1)\sigma_j^{\alpha\alpha'}\sigma_j^{\gamma\gamma'}u_2^2 \
,\label{spinAverage}
\end{align}
is carried out
employing $ \overline{ S_i}=0$ and $ \overline{ S_i
S_j}=\delta_{ij}\;S(S+1)/3\;$ (where $i,j=x,y,z$).

Following Ref.~\onlinecite{AG} we assume that
$\;K_{\omega\alpha\gamma}=\epsilon_{\alpha\gamma} K_{\omega}$,
and then using $\sigma_j^{\alpha\alpha'}\sigma_j^{\gamma\gamma'}
\epsilon_{\alpha'\gamma'}= -3\epsilon_{\alpha\gamma}\;$ we
obtain
\begin{align}\label{Pi_and_K}
\Pi({\bf q},\nu)=&\sum_{\omega}K_{\omega}({\bf q},\nu)\;,
\nonumber\\
K_{\omega}({\bf q},\nu)&=\left[ 1+ (2\pi {\cal N}(0)\tau_-)^{-1}
K_{\omega}({\bf q},\nu)\right]\nonumber\\& \times \sum_{\bf k}
\bar G({\bf k}+{\bf q},\omega+\nu) \bar G(-{\bf k},-\omega) \ ,
\end{align}
where the averaged Green function is
\begin{align}\label{average_G}
\bar G({\bf p},\omega)= [i\omega-({\bf p}^2/2m-\mu)+i{\rm
sgn}(\omega)/2\tau_+]^{-1}\ .
\end{align}
(The spin indices are suppressed since $\bar G$ is independent
of them.) In Eqs.~(\ref{Pi_and_K}) and (\ref{average_G}),
\begin{align}
\frac{1}{\tau_\pm}=2\pi {\cal N}(0)N_i (u_1^2\pm S(S+1)u_2^2)\
,
\end{align}
where ${\cal N}(0)$ is the extensive density of states at the
Fermi level. (Note that $\tau_+$ is the elastic mean free time.)
Using \equref{average_G} to calculate the sum over {\bf k} in
\equref{Pi_and_K} yields
\begin{align}\label{sumGG}
&\sum_{{\bf k}'} \bar G({\bf k}'+{\bf q},\omega+\nu) \bar
G(-{\bf k}',-\omega)= 2\pi {\cal N}(0)\tau_+\nonumber\\
&\times\theta[\omega(\omega+\nu)] (1-\tau_+|2\omega+\nu|- D{\bf
q}^2\tau_+ )\ .
\end{align}
Upon inserting this expression into \equref{Pi_and_K} and
solving it one finds
\begin{align}\label{K}
&K_{\omega}({\bf q},\nu)= 2\pi {\cal
N}(0)\theta[\omega(\omega+\nu)]\nonumber\\ &\times (D{\bf
q}^2+|2\omega+\nu|+2/\tau_s)^{-1}\ ,
\end{align}
where $1/\tau_s$ is the pair-breaking rate
\begin{align}
\frac{1}{\tau_{s}}= 2\pi {\cal N}(0)N_i S(S+1)u_2^2\ .
\end{align}
When $\tau_+\simeq\tau_-$ most of the disorder is due to the
non-magnetic part. This, together with the assumption\cite{AGD}
$\{|2\omega+\nu|,D{\bf q}^2\}\ll 1/\tau_+$ were used in
obtaining \equref{K}.

The summation in \equref{Pi_and_K} over the Matsubara
frequencies can be written explicitly as
\begin{align}
\label{eq:Pi}
\frac{T}{{\cal N}(0)}\Pi({\bf q},\nu) =\sum_{
\tilde n=0}^{\infty}\left[ \tilde n+\frac{1}{2}+
\frac{|\nu|+2/\tau_s+D{\bf q}^2}{4\pi T}\right]^{-1}\ .
\end{align}
Note that \equref{eq:Pi} includes also the negative Matsubara
frequencies. This sum does not converge and therefore a cutoff
is required. The cutoff frequency on the attractive interaction
is the Debye frequency $\omega_D$, and consequently the sum is
terminated at $\tilde n =\omega_D/2\pi T$. As a result, the
polarization is given by
\begin{align}\label{Pi:final}
\frac{T}{{\cal N}(0)}&\Pi({\bf q},\nu)= \Psi \Bigl
(\frac{1}{2}+\frac{
\omega_D}{2\pi T}+\frac{|\nu|+2/\tau_s+D{\bf q}^2}{4\pi T}\Bigr )\nonumber\\
&-\Psi \Bigl (\frac{1}{2}+\frac{|\nu|+2/\tau_s+D{\bf q}^2}{4\pi
T}\Bigr )\ ,
\end{align}
where $\Psi$ is the digamma function.

We next express the polarization in terms of the bare transition
temperature of the system. This is  the temperature at which
${\cal Z}/{\cal Z}_0$ diverges for $|\nu|=0$ and the smallest
possible $|\bf q|$, in the absence of the pair breakers and the
magnetic flux,
\begin{align}\label{Tc0}
\frac{V}{g{\cal N}(0)}=\Psi\Bigl
(\frac{1}{2}+\frac{\omega_D}{2\pi T_c^0} \Bigr )- \Psi\Bigl
(\frac{1}{2}\Bigr )\ .
\end{align}
Since $\omega_D\gg \{T_c^0,T\}$ we may use the asymptotic
expansion of the digamma function,
\begin{align}
\label{derivation:psi:asymptotic} \Psi(x\gg 1)\simeq \ln(x)\ ,
\end{align}
to obtain
\begin{align}
{\cal Z}&={\cal Z}_0\prod_{{\bf q},\nu} \Bigl ( \frac{V}{g{\cal
N}(0)}\;
 \Bigl [ \ln\Bigl
(\frac{T}{T_c^0}\Bigr )\nonumber\\
&+ \Psi\Bigl (\frac{1}{2}+ \frac{|\nu|+2/\tau_s+D{\bf q}^2}{4\pi
T}\Bigr )-\Psi\Bigl (\frac{1}{2}\Bigr )\Bigr ]^{-1} \Bigr ) \
.\label{SOF}
\end{align}
The effect of the pair
breakers is represented by the term $2/\tau_s$ in the argument
of the digamma functions.

As is mentioned above, the persistent current is given by
\begin{align}
I=(e/2\pi)\;\partial T\ln{\cal Z}/\partial \phi \ .\label{DERZ}
\end{align}
The flux enters the expression for ${\cal Z}$ through the
longitudinal components of the momenta, see
\equref{momentaBose}. In our ring geometry, only momenta of zero
transverse components contribute significantly to the current,
since the contribution of momenta of higher transverse
components can be shown to decay exponentially, as a function of
the ratio of $L$ and the transverse dimension (e.g., the height)
of the ring.

As is seen in Eqs.~(\ref{SOF}) and (\ref{DERZ}), the PC consists of
two parts. The first arises from differentiating ${\cal Z}_{0}$ and
is the ensemble averaged PC of non-interacting, grand-canonical,
normal metal rings. \cite{CGR} This contribution is much too small
to account for the measured amplitude of the $h/2e$ harmonic [see
\secref{section:Introduction}], and therefore will be omitted in the
following. The other part of the PC comes from the free energy due
to the superconducting fluctuations,
\begin{align}
&I=-2eE_{c}\sum_{n,\nu} \frac{(n+2\phi) \Psi '(\tilde F(n, \nu
))}{\ln(T/T_c^0)+\Psi(\tilde F(n, \nu ))-\Psi(\frac{1}{2})} \
,\label{PC_before Poisson}\end{align} where we have introduced
the function
\begin{align}
\tilde F(n,\nu )=\frac{1}{2}+\frac{|\nu|+2/\tau_s}{4\pi
T}+\frac{\pi E_c }{T}(n+2\phi)^2\ .
\end{align}
In particular, one notes the $h/2e$ periodicity in the flux.
Indeed, upon employing the Poisson summation formula
\begin{align}
&I=-8eE_{c}\sum_{m=1}^{\infty} \frac{\sin(4\pi
m\phi)}{m^2}\nonumber\\
&\times\sum_\nu \int_0^\infty dx\frac{x \sin(2\pi x) \Psi '(F(x,
\nu ))}{\ln(T/T_c^0)+\Psi(F(x, \nu ))-\Psi(\frac{1}{2})} \
,\label{MAIN}
\end{align}
where
\begin{align}
F(x,\nu )=\frac{1}{2}+\frac{|\nu|+2/\tau_s}{4\pi T}+\frac{\pi
E_c x^2}{m^2 T}\ .\label{FF}
\end{align}
Clearly, the fluctuation-induced PC decreases as the
pair-breaking strength increases. Our central result is that
this decrease may be far less than the one caused in the
transition temperature.

In order to compare the dependence of the PC and of the
transition temperature on the pair-breaking strength, we use the
expression \cite{AG} for the transition temperature in the
presence of both pair breakers and magnetic flux
\begin{align}
&\ln\Bigl (\frac{T_c}{T_c^0}\Bigr )\nonumber \\&+ \Psi\Bigl
(\frac{1}{2}+\frac{4\pi E_c\phi^2}{T_c}+ \frac{1}{2\pi
T_c\tau_s}\Bigr )-\Psi\Bigl (\frac{1}{2}\Bigr )=0\ .\label{Tc}
\end{align}
Here $\phi $ is in the range $-1/2, 1/2$, modulo unity.
\cite{ANOTHERCOM} We plot in \figref{1} the amplitude of the
$h/2e$ harmonic of the PC, as well as the transition temperature
(in the absence of the flux) as functions of the pair-breaking
strength, using the dimensionless parameter
\begin{align}\label{derivation:s}
s=1/\pi T_c^0 \tau_s \ .
\end{align}
The transition temperature is reduced due to pair breaking and
vanishes at $s=1/2\gamma_E$, where $\gamma_E$ is the Euler
constant. In contrast, for $E_c\gg 1/\tau_s$ the PC is hardly
affected for these values of pair-breaking strengths.

\begin{figure}[thd]
\begin{center}
\includegraphics[width=8.6cm]{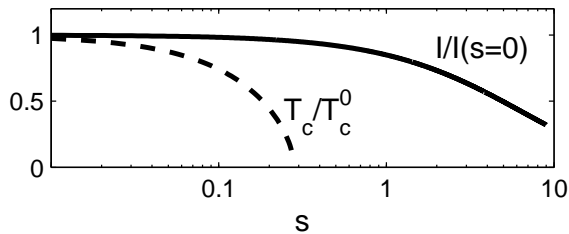}
\end{center}\vspace{-0.5cm}
\caption{The $h/2e$ harmonic (full line) and $T_c/T_c^0$ (dashed
line) as functions of the pair-breaking strength, displayed on a
logarithmic scale. The current, in units of $I(s=0)$, is plotted
for $T=E_c$ and $T_c^0=0.1 E_c$. The PC reduction at $s=10$
corresponds to $1/\tau_s=\pi E_c$.} \label{fig1}
\end{figure}

\begin{figure}[tbp]
\begin{center}
\includegraphics[width=8.6cm]{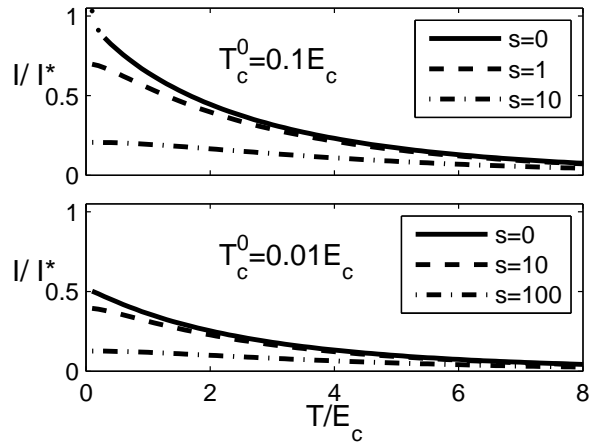}
\end{center}\vspace{-0.5cm}
\caption{The amplitude of the $h/2e$ harmonic in units of $I^*=-
e E_c$, as a function of the temperature, for two values of
$T_c^0/E_c$ and several values of $s$. Note that the $s=0$ curve
in the upper panel is valid only for $T/T_c\geq 1+Gi$, where
$Gi$ is the Ginzburg parameter.} \label{fig2}
\end{figure}

Figure~\ref{fig2} portrays the PC plotted by numerically
evaluating Eq.~(\ref{MAIN}). In each of the panels the upper
curve is drawn for $s=0$, while the second curve corresponds to
a pair-breaking strength [see \equref{Tc} and \figref{1}] which
is large enough to destroy $T_c$. Nonetheless, the PC is hardly
affected as long as $L_{s}\gtrsim L$ [see Eqs.~(\ref{2}) and
(\ref{introduction:LsXi0})]. The considerably-reduced PC due to
a small $L_s$ is presented by the dash-dotted curves, which
correspond to $L_s\simeq 0.5 L$. The effect of the temperature
on the magnitude of the PC is manifested by its dependence of
the ratio $L/L_T$, where $L_T$ is the thermal length,
\begin{align}\label{L_T}
L_T=\sqrt{D/T}\ ,
\end{align}
or equivalently the ratio $T/E_c$, see \figref{2}.

\section{The dominant fluctuations}\label{section:DominantFluctuations}

Our result for the PC [see Eq.~(\ref{PC_before Poisson})]
consists of infinite sums over the frequencies and over the
momenta. One naturally asks oneself whether the characteristic
features of the expression are not given by the first few
members of each sum, notably the static, $\nu=0$, regime. It
turns out that this is not the case over most of the relevant
range: to obtain the correct magnitude of the
fluctuation-induced PC, numerous frequencies and momenta are
required.

\begin{figure}[thd]
\begin{center}
\includegraphics[width=8.6cm,angle=0]{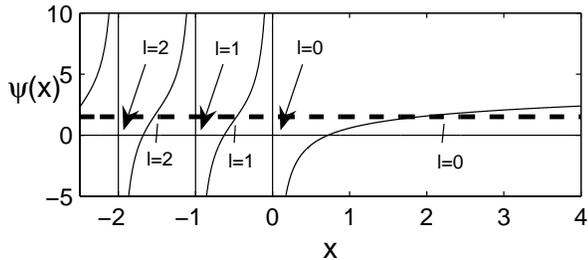}
\end{center}\vspace{-0.5cm}
\caption{The digamma function (solid line) and
$\ln(T^{0}_{c}/4\gamma_{E}T)$ for $T_c^0/T=0.6$ (dashed line).
The first three solutions $F^{\ell}_{\rm zero}$ of
\equref{eq:F_l_zero} are marked on the x axis with their indices
indicated below it. The first values of the set $F^{\ell}_{\rm
pole}$, \equref{eq:F_l_pole}, are marked by arrows.}
\label{fig:digamma}
\end{figure}

In order to study this aspect, it is convenient to express the
PC in a form which is more amenable to numerical computations.
To this end we write \equref{MAIN} as
\begin{align}
&I=\frac{2ieT}{\pi}\sum_{m=1}^{\infty}\sin (4\pi m\phi)\sum_{\nu}\int_{-\infty}^{\infty}dx e^{2\pi ix}\nonumber\\
&\times \frac{d}{dx}\ln\left[\Psi
(F(x,\nu))-\ln(T^{0}_{c}/4\gamma_{E}T)\right]\ ,\label{MAINSIM}
\end{align}
where the function $F$ is given in \equref{FF}. The
$x$-integration is carried out by closing the integral in the
upper half of the complex plane. Two sets of simple poles can be
identified in the integrand of \equref{MAINSIM}. These sets
result from (a) the zeros and (b) the poles of the argument of
the logarithm.\cite{COM4} The first set of poles, denoted by
$x^{\ell}_{\rm zero}$, is given by
\begin{align}
\Psi(F^{\ell}_{\rm zero})=\ln(T^{0}_{c}/4\gamma_{E}T)\ .
\label{eq:F_l_zero}
\end{align}
The second set consists of the poles of the digamma function.
These are denoted by $x^{\ell}_{\rm pole}$, and are obtained
from the relation
\begin{align}
F^{\ell}_{\rm pole}=-\ell \ , \ \ \ell =0,1,2,\ldots\ .
\label{eq:F_l_pole}
\end{align}
The index $\ell$ runs over the poles in each set. The two sets
of $F^{\ell}_{\rm pole/zero}$ given by Eqs.~(\ref{eq:F_l_zero})
and (\ref{eq:F_l_pole}), are shown in \figref{fig:digamma}.

Performing the Cauchy integration, the current takes the
form\cite{COM5}
\begin{align}
I=& -4e T\sum_{m=1}^{\infty} \sin(4\pi m\phi)\nonumber\\
&\times \sum_\nu\sum_{\ell =0}^\infty \Bigl [ \exp(2\pi i
x_{\textrm{zero}}^{\ell})
 - \exp(2\pi i x_{\textrm{pole}}^{\ell}) \Bigr ] \ .\label{eq:contour}
\end{align}
Here $x^{\ell}_{\rm pole/zero}$ depends on the Matsubara
frequency and the harmonic index $m$,
\begin{align}
& x_{\rm pole/zero}^{\ell}\nonumber\\&=im\sqrt{\frac{T}{2\pi
E_c}}\Bigl [1+\frac{|\nu|+2/\tau_s}{2\pi T }-2F_{\rm
pole/zero}^{\ell}\Bigr ]^{1/2}\ . \label{eq:expansion:x}
\end{align}
Note that all the exponents ($2\pi i x$) in the two series in
\equref{eq:contour} are negative, and their absolute value
increases with increasing $\nu,l$ or $m$. As can be seen from
\figref{fig:digamma}, for each pair of poles $F^{\ell}_{\rm
zero}>F^{\ell}_{\rm pole}$, and consequently $|x^{\ell}_{\rm
zero}|<|x^{\ell}_{\rm pole}|$. This ensures that the term in the
square brackets of \equref{eq:contour} is positive, and hence
the response of the ring to a small flux is diamagnetic, as it
should be.

\subsection{The dominant imaginary time fluctuations}
\label{subsec:matsubaraFreq}

The dominant terms in \equref{eq:contour} are those for which
the absolute value of $x$ is smaller than unity, but if the
absolute values of all $x$ are larger than one only the smallest
[$x_{\rm zero}^{\ell=0}(\nu=0,m=1)$] is the dominant one. The
absolute value of the exponents (which are given by $2\pi
|x_{{\rm zero}/{\rm pole}}|$) is at least $(|\nu|/ E_c)^{1/2}$.
Thus, the frequencies that contribute mostly to the current are
those for which $|\nu|\lesssim 10 E_c$. The proportionality
factor, of order $10$, had been determined numerically and
resulted from the square-root structure of the exponents, see
\equref{eq:expansion:x}. At high temperatures $T>E_c$ the system
is dominated by the classical fluctuations - namely, by the
first (lowest energy), $\nu=0$, Matsubara frequency. The effect
of the quantum fluctuations for which $\nu\neq 0$ increases as
the temperature decreases. This tendency has an exception in two
cases. First, for very strong pair breaking
$1/\tau_s>\{T,E_c,T^2/E_c\}$ the significant quantum
fluctuations that have a dominant contribution to the PC are
bounded by $|\nu|<\sqrt{E_c/\tau_s} $. Second, in the case of
small or zero pair breaking when $T\rightarrow T_c$, only
$\nu=0$ is the dominant frequency.\cite{LV}

When $T_c$ is finite, the $n=\nu=0$ pole of the partition function,
\equref{ZwithPi}, is the most dominant one as $T\rightarrow T_c$.
Consequently, in this low-temperature regime physical properties,
including the PC, are determined only by the $\nu=0$ fluctuations,
pertaining to the static Ginzburg-Landau free energy. We find
however, that in the case of a vanishing $T_c$, quantum
fluctuations, for which $\nu\neq 0$ have a significant contribution
to the PC at low temperatures. Indeed, the quantum fluctuations of a
system with no magnetic impurities and for which
$|\phi|^2>T_c^0/(16\pi \gamma_E E_c)$ have been recently invoked in
the context of the ``strong" Little-Parks oscillations, see Ref.
\onlinecite{SO}.

\subsection{The dominant spatial fluctuations}

\begin{figure}[thd]
\begin{center}
\includegraphics[width=8.6cm,angle=0]{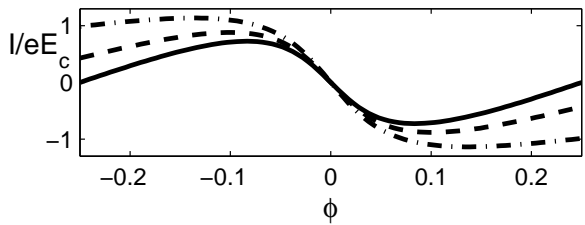}
\end{center}\vspace{-0.5cm}
\caption{The PC as computed from \equref{PC_before Poisson},
with the summation over $n$ cut at $1000,3,1$ (solid, dashed and
dash-dotted curves, respectively). The plots are for
$T=T_c^0=0.1 E_c$ and $s=1$.} \label{fig:qI_of_phi}
\end{figure}

High Matsubara frequencies involve many spatial frequencies ${\bf
q}$. Thus, at low temperatures and for a vanishing $T_c$, many wave
vectors contribute to the PC. We have estimated numerically their
number by comparing the PC computed with a relatively small number
of frequencies and wave vectors with the exact result
\equref{eq:contour} for $T=T_c^0=0.1E_c$ and $s=1$. In this case we
have found that $\sim 100$ Matsubara frequencies are required. The
highest momenta, \equref{momentaBose}, that contribute significantly
are given by $|n|\sim (1,5,100,1000)$ for the frequencies $\nu/(2\pi
T)=(0,5,10,100)$, respectively. Figure \ref{fig:qI_of_phi} shows the
PC as computed from \equref{PC_before Poisson} for different maximal
$|{\bf q}|$ values and without limiting the range of $\nu$. It is
thus seen that in the whole range of $\phi$ the persistent current
is not mainly determined by the lowest momenta, even when the size
of the system $L$ is smaller than the thermal length $L_T$,
\equref{L_T}. This is different than the situation in the
calculations of other properties (for example, weak-localization
corrections \cite{0D} to the conductivity), in which $L\ll L_T$ is
taken as a sufficient condition for using only $q=0$. We point out
however that the slope of the PC at $\phi=0$ appears to be
describable using the smallest wave number only.

\subsection{The dominant harmonics}

\begin{figure}[thd]
\begin{center}
\includegraphics[width=8.6cm,angle=0]{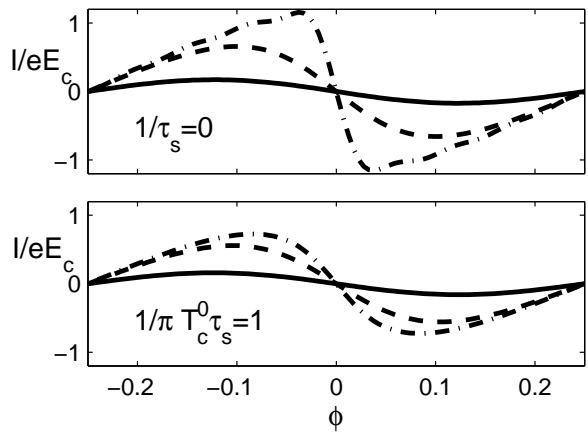}
\end{center}\vspace{-0.5cm}
\caption{The current, in units of $eE_c$, as a function of the
flux $\phi$, for $T_c^0/E_c=0.1\;$ and for several temperatures,
$T/E_c=5,1$ and $0.15$ in the solid, dashed and dash-dotted
curves respectively. In the lower panel the dash-dotted curve
corresponds to $T/E_c=0.1$. For $s=0$ the current attains the
sawtooth form (upper panel) which is lost for $s=1$ (lower
panel). }\label{fig:m_harmonics}
\end{figure}

Examining the series in
\equref{eq:contour} one can see that the maximal harmonic of the
flux, $m_{\max}$, that has still a significant contribution to the
current is given by $\min\{ \sqrt{E_c/T}, \sqrt{E_c\tau_s}\}$ or
by one if the first two values are smaller than unity. This
condition can be expressed in terms of lengths by
\begin{align}\label{max_m}
m_{\max}=\min\{ L_s/L,\ L_T/L \}\ ,\ \textrm{ or 1}.
\end{align}
The upper limit on the harmonics results from the fact that the
$m$'th harmonic is associated with paths that encircle the ring
(coherently) $m$ times and hence their length is at least $mL$.
\cite{Argaman} The sinusoidal shape $I\propto \sin(4\pi \phi)$ at
high temperatures is modified due to higher harmonics as the
temperature decreases. In the absence of magnetic impurities (upper
panel in \figref{fig:m_harmonics}) the low-temperature current as a
function of the flux attains a sawtooth shape. Such a behavior is
predicted also for the equilibrium PC in superconductors at zero
temperature \cite{book} and for the persistent current in a clean
system of non-interacting
electrons.\cite{Cheung_Gefen_Riedel_Shih_1988} In the presence of
pair-breakers the upper bound on the harmonics \equref{max_m}
prevents the current from reaching the sharp sawtooth shape. This
suggests, in principle, a way to experimentally confirm the role of
pair breaking for this problem. In the lower panel of
\figref{fig:m_harmonics} the current of a system with $L\simeq L_s$
is plotted for several temperatures. At temperatures below $0.1 E_c$
the shape of the current does not change anymore.

\section{The temperature dependence}
\label{section:limits}

Here we study the PC in the low and high-temperature regimes. In
particular we find that the PC decays exponentially as the length
of the ring exceeds the thermal length $L_{T}$ or the magnetic
impurity scattering length $L_{s}$, whichever is shorter.

\subsection{High-temperature regime, $T\gg \max\{1/\tau_s,T_c^0,E_c \}$}

When the temperature is much higher than all relevant energy
scales, i.e., $T\gg \max\{1/\tau_s,T_c^0,E_c \}$, the leading
contribution to the double sum in \equref{eq:contour} comes
solely from the first pole $x^{\ell =0}_{\rm zero}$ of the lowest
Matsubara frequency, $\nu =0$ [see \equref{eq:expansion:x}]. In
this temperature range the $h/2e$ harmonic, corresponding to $m=1$, is the dominant
one.

\begin{figure}[thd]
\begin{center}
\includegraphics[width=8.6cm]{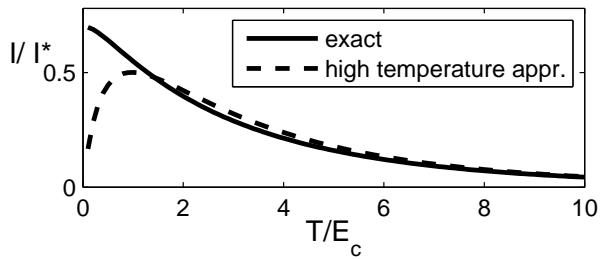}
\end{center}\vspace{-0.6cm}
\caption{The amplitude of the $h/2e$ harmonic is plotted in
units of $I^*=- e E_c$ as a function of the temperature, for
$T_c^0/E_c=0.1$ and $s=1$. The exact results can be approximated
by \equref{eq:very_high_T} for $T\gg E_c$. } \label{fig:highT}
\end{figure}

As the temperature increases, the horizontal line in
\figref{fig:digamma} representing $\ln(T_c^0/4\gamma_E T)$ moves
further down, so that $F_{\textrm{zero}}^{\ell=0}$ approaches
zero. We use the expansion of the digamma function for small
arguments in \equref{eq:F_l_zero} and obtain
\begin{align}
F_{\textrm{zero}}^{\ell=0}= \left[ \ln(T_c^0/4\gamma_E^2 T)
\right]^{-1}\ .
\end{align}
Upon substituting this result in the dominant term of
\equref{eq:contour}, we obtain the current in the form
\begin{align}
&I\simeq -4e T \sin(4\pi \phi) \nonumber\\
&\times \exp\Bigl (- \frac{L}{L_T}\Bigl [2\pi+\frac{2L^{2}_{T}}{
L^{2}_{s}}-\frac{4}{\pi\ln(4\gamma_E^2 T/T_c^0)}\Bigr
]^{1/2}\Bigr )\ . \label{eq:very_high_T}
\end{align}
We compare the full result, \equref{eq:contour}, with the
high-temperature approximation \equref{eq:very_high_T} in
\figref{fig:highT}. The difference between the contributions of
the first $x^{\ell =0}_{\rm zero}(\nu=0)$ and the second
$x^{\ell =0}_{\rm pole}(\nu=0)$ poles to the PC is the absence
of the third term, which includes a logarithm [see
\equref{eq:very_high_T}], in the exponent of the latter.
Therefore, this approximation improves as $T_c^0$ increases.

\subsection{Low-temperature regime, $T_c\ll T\ll \{1/\tau_s,E_{c}\}$}
In the low temperature regime the argument ($F$) of the digamma
function and its derivative is much larger than unity [see
\equref{FF}], so that we can use their asymptotic expansions
$\ln(F)$ and $1/F$, respectively. Substituting these approximations
in \equref{MAIN} gives
\begin{align}
&I=-\frac{8}{\pi}\;eT\sum_{m=1}^{\infty} \sin(4\pi
m\phi)\nonumber\\
&\times\sum_\nu \int_0^\infty \frac{x \sin(2\pi x)dx
}{\ln\left[\frac{4\pi \gamma_E
E_c}{T_c^0}(x^2+a_{m,\nu})\right](x^2+a_{m,\nu})} \ ,
\label{LowT:MAIN}
\end{align}
where $a_{m,\nu}=m^2(|\nu|+2/\tau_s+2\pi T)/(4\pi^2 E_c)$. For
$T\gg T_c$ the denominator in \equref{LowT:MAIN} does not
vanish. Then the term $x^2+a_{m,\nu}$ in the logarithm in
\equref{LowT:MAIN} can be replaced by $\alpha a_{m,\nu}$, with,
say, $1<\alpha<3$. Consequently,
\begin{align}\label{eq:lowT1}
I\simeq &-4 e T \sum_m \sin(4\pi m \phi)\sum_\nu
e^{-m\sqrt{\frac{ 2\pi T}{E_c}}
\sqrt{1+\frac{|\nu|+2/\tau_s}{2\pi T}}} \nonumber
\\ & / \ln\left[\frac{2\gamma_E\alpha
T}{T_c^0}\left(1+\frac{|\nu|+2/\tau_s}{2\pi T}\right)\right] \ .
\end{align}
Since $T\ll E_c$ the summation over $\nu$ can be replaced by an
integration. Approximating again the logarithm by its value at
the dominant $\nu$ of the integration, yields
\begin{align}\label{eq:lowT2}
I\simeq &-\frac{8}{\pi}\; e E_c \sum_m \frac{\sin(4\pi m
\phi)}{m^2} \left[ 1+m\sqrt{\frac{2\pi L^2}{ L_T^2}+\frac{2L^2}{
L_s^2}} \right]\nonumber \\ & \times e^{-m\sqrt{\frac{2\pi L^2}{
L_T^2}+\frac{2L^2}{ L_s^2}}} / \ln\left[\frac{\gamma_E\alpha
E_c}{\pi T_c^0 m^2} \bar z\right]\ ,
\end{align}
where $\bar z=\max\{ 1,2m^2/\tau_s E_c\}$.

\begin{figure}[thd]
\begin{center}
\includegraphics[width=8.6cm,angle=0]{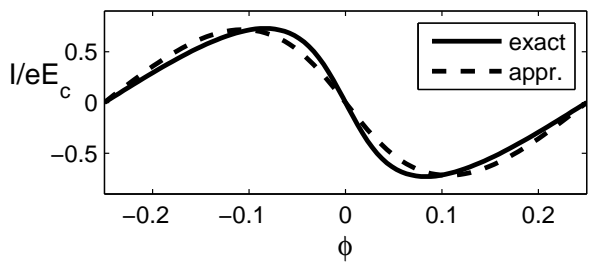}
\end{center}\vspace{-0.5cm}
\caption{The current in units of $eE_c$ as a function of the
magnetic flux $\phi$, plotted for $T=0.1T_c^0=0.01E_c$ and $s=1$.
The low-temperature approximation \equref{eq:lowT2} is compared
with the exact result \equref{eq:contour}. We take $\alpha=3$ in
the logarithm of \equref{eq:lowT2}.} \label{fig:limit:Iphi}
\end{figure}

We compare in \figref{fig:limit:Iphi} the low-temperature
approximation, \equref{eq:lowT2}, with the full result,
Eq.~(\ref{eq:contour}). As one can see from this comparison, the
flux dependence of the PC as well as its amplitude are well
approximated by \equref{eq:lowT2}.

\section{Renormalization of the effective interaction}
\label{section:renormalization} In this section we calculate the
PC to first order in the interaction, in order to see whether it
suffices to explain our full result. To first order in the
interaction, the contribution of superconducting fluctuations to
the free energy [see \equref{ZwithPi}] is
\begin{align}\label{conclusions:Ffirst}
\Delta \Omega=-(gT^2/V)\sum_{{\bf q},\nu}\Pi({\bf q},\nu)\ .
\end{align}
The PC resulting from \equref{conclusions:Ffirst} has the same form
as \equref{PC_before Poisson}, except that the denominator in the
latter is replaced by the bare interaction $g{\cal N}(0)/V$. Had we
tried to to fit the experimental data of Refs. \onlinecite {LDDB}
and \onlinecite{JMKW} using \equref{conclusions:Ffirst}, we should
have taken the implausible ratio $E_c\sim 0.1 \omega_D$ [see
\equref{Tc0}]. This first-order approximation fails because of
screening effects, which increase the magnitude of the effective
attractive interaction as the temperature decreases. Very roughly,
the renormalization of a dimensionless interaction $\lambda$, from a
higher frequency scale $\omega_>$ to a lower one, $\omega_<$, is
given by\cite{MA}
\begin{align}
\lambda (\omega_<) = \left[\lambda^{-1}( \omega
_>)-\ln\left(\frac{\omega_>}{\omega_<}\right)\right]^{-1}\ .
\end{align}
For attractive interactions $\lambda$ is positive and the high
frequency scale is $\omega_D$. At $T=T^0_c$ and $1/\tau_s=0$,
the attractive interaction should diverge. Using this to
eliminate $\lambda (\omega_D$) ($\equiv g {\cal N}(0)/V$), we
obtain that for $T^0_c \lesssim \omega \ll \omega_D,$
\begin{align}\label{conclusion:lambdaRough}
\lambda(\omega) \backsim 1/\ln (\omega /T_c^0)\ .
\end{align}
Replacing in the first-order approximation for the current the
bare interaction by the effective one,
\equref{conclusion:lambdaRough}, gives
\begin{align}\label{conclusions:PCfirst}
&I_{\textrm{1st}}=-\frac{8eE_{c}}{\ln(\omega/T_c^0)}\sum_{m=1}^{\infty}
\frac{\sin(4\pi
m\phi)}{m^2}\nonumber\\
&\times\sum_\nu \int_0^\infty dx x \sin(2\pi x) \Psi '(F(x, \nu
))\ .
\end{align}

\begin{figure}[thd]
\begin{center}
\includegraphics[width=8.6cm,angle=0]{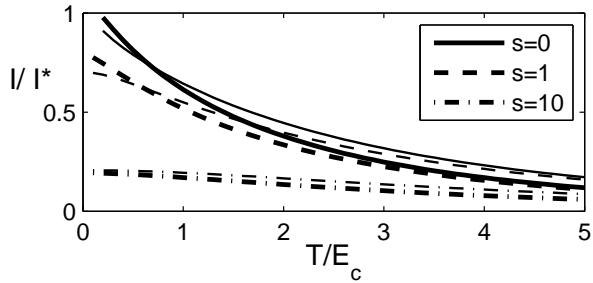}
\end{center}\vspace{-0.5cm}
\caption{The first-order approximation for the $h/2e$ harmonic
of the current \equref{conclusions:PCfirst} (bold lines) is
compared with the exact result (thin lines). Here
$T_c^0=0.1E_c$. In drawing the former, we have used the simplest
expression for the cutoff $\omega=T+E_c+1/\tau_s$ .}
\label{fig:first}
\end{figure}

The effective interaction is renormalized upwards with
decreasing energy and, for the bulk and no pair breaking, it
blows up at $T_c^0$. For $1/\tau_s>T_c^0$, this renormalization
stops at $1/\tau_s$ and $T_c$ disappears. In the mesoscopic
range, the Thouless energy, $E_c$, becomes a relevant scale and
it may be expected (as is borne out by our results) that the PC
at low temperatures is determined by the interaction on that
scale, as long as $E_c \gtrsim 1/\tau_s$. Once $1/\tau_s\gtrsim
E_c$, we expect the renormalization to ``stop at $1/\tau_s$" and
the PC to be depressed. Thus, the relevant range for our
considerations is $T_c^0\lesssim 1/\tau_s\lesssim E_c$. Using
these bounds on the energy scale of the renormalized interaction
in the first order calculation \equref{conclusions:PCfirst},
gives a good agreement with our result \equref{eq:contour}. In
\figref{fig:first} we plot the amplitude of the $h/2e$ harmonic
as a function of $T/E_c$, calculated from the full expression
\equref{MAIN} (thin curves) and from the first-order
approximation Eq.~(\ref{conclusions:PCfirst}) (bold curves).
The plotted curves are for $T_c^0=0.1E_c$.

A more precise expression for the renormalized attractive
interaction depends on ${\bf q},\nu$ of the order-parameter
fluctuation. The renormalized attractive interaction $\lambda({\bf
q},\nu)$, obtained from an infinite series of diagrams containing
Cooperon contributions, is given by \cite{AA}
\begin{align}\label{conclusion:lambdaExact}
&\lambda({\bf q},\nu)=\Big[ \lambda^{-1}(\omega_D)-\frac{T}{{\cal
N}(0)}\Pi({\bf q},\nu)\Big ]^{-1}\ .
\end{align}
Upon substituting \equref{Pi:final} in
\equref{conclusion:lambdaExact} one can identify $\lambda({\bf
q},\nu)$ from our result, e.g., by comparing
\equref{conclusions:PCfirst} with \equref{PC_before Poisson}.

\section{Comparison with experiments}\label{section:compare_experiments}
Theoretically, only static magnetic fields have been considered
here. However, experiments have been carried out with an ac magnetic
field. In the experiments on copper\cite{LDDB} and gold,\cite{JMKW}
the sweeping frequencies of the magnetic field were very low ($0.3$
Hz and $2$Hz respectively). Thus, one expects that the measured PC
could be explained using a theory for a static magnetic field. In
the experiment on silver, on the other hand, a very high sweeping
frequency of the magnetic field was used ($217$ MHz). It is
plausible that in order to explain the results of
Ref.~\onlinecite{DBRBM} one may not confine oneself to a static
magnetic field. We therefore do not attempt to explain the
experiment of Ref. \onlinecite{DBRBM}.

Here we explain the $h/2e$ signal observed in copper\cite{LDDB}
and gold\cite{JMKW} using our result \equref{eq:contour}. In the
left six columns of \tableref{table:experiments} we summarize
the experimental parameters for the $(h/2e)$-periodic
signal.\cite{COM6}

\begin{widetext}
\begin{center}
\begin{table}[h]
\begin{tabular}{|p{1.4cm}|p{1.4cm}|p{1.4cm}
|p{1.4cm}|p{1.4cm}|p{2.2cm}||p{3.0cm}|} \hline & $E_c$ & $T$
&$I/eE_c$ & $L$ & $L_\varphi$
& min $T_c^0$ \\
\hline \hline Copper\cite{LDDB} & $15$ mK& $7$ mK&$1$&$2.2\;\mu
$m &$2\;\mu \textrm{m} (1.5\textrm{ K})$ & a few mK\\
\hline Gold\cite{JMKW} & $4.9$ mK & $5.5$ mK&$0.65$&$8.0\;\mu $m
&$16\;\mu \textrm{m} (0.5\textrm{ K})$& a fraction of a mK \\
\hline
\end{tabular}
\caption{Experimental parameters in the left six columns. The
magnitude of the $h/2e$ periodic current (column $4$) is given for
the lowest temperature (column $3$) reached in the experiment. The
dephasing length $L_\varphi$ is given together with the temperature
at which it was measured. The last column is our estimate for a
lower bound on $T_c^0$ according to Eq.~(\ref{eq:contour}), see also
\figref{fig:Tc0_over_Ec_versus_Ls_over_L}.}
\label{table:experiments}
\end{table}
\end{center}
\end{widetext}

\begin{figure}[t]
\begin{center}
\includegraphics[width=8.6cm,angle=0]{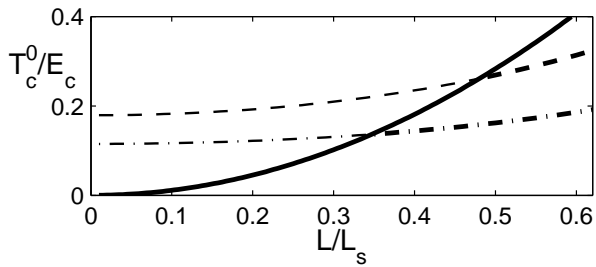}
\end{center}\vspace{-0.5cm}
\caption{The bare transition temperatures corresponding to the
measured PC as a function of $L/L_s$. The dashed and dash-dotted
curves correspond to the PC measured in copper and gold,
respectively. The solid curve gives the maximal possible $T_c^0$
satisfying $T_c=0$. } \label{fig:Tc0_over_Ec_versus_Ls_over_L}
\end{figure}

The metals used in the experiments are not superconductors at any
measured temperature in their bulk form. Therefore, it is not
possible to obtain theoretically a large enough PC (to match the
measurments\cite{LDDB,JMKW}) due to the attractive interaction
\textit{without} pair breaking: the required $T_c\sim 1$~mK are too
high to be considered as realistic. We suggest that the bare
transition temperature may indeed be of the order of a mK, but the
transition temperature of the real material is considerably reduced due
to pair breakers. Together with this assumption, the necessary
condition to fit the experiments is $1/\tau_s \gtrsim \pi
T_c^0/2\gamma_E$ so that $T_c$ vanishes or is very strongly
depressed,\cite{AG} see \equref{Tc}. This condition can also be
written as
\begin{align}\label{fittingExperiments:condition}
\frac{T_c^0}{E_c}\lesssim
\frac{2\gamma_E}{\pi}\left(\frac{L}{L_s}\right)^2\ .
\end{align}
Note that we need $L_s\gtrsim L$ in order not to depress the PC
[\equref{eq:contour}]. The upper limit on $T_c^0$, corresponding
to the equality in \equref{fittingExperiments:condition}, is
given by the solid line in
\figref{fig:Tc0_over_Ec_versus_Ls_over_L}. The values for
$T_c^0/E_c$ that correspond to a vanishing $T_c$ are in the
region below this line. In the dashed and dash-dotted curves in
\figref{fig:Tc0_over_Ec_versus_Ls_over_L} different values of
$L/L_s$ are matched with an appropriate $T_c^0/E_c$ so that the
measured values in columns $2-4$ of \tableref{table:experiments}
remain the same.

The monotonically increasing shape of the curves in
\figref{fig:Tc0_over_Ec_versus_Ls_over_L} results from the fact
that higher values of $T_c^0/E_c$ are required to describe the
experiments as $L/L_s$ increases. The minimal $T_c^0$'s
correspond to the points where the dashed and the dash-dotted
lines cross the solid line. In this way we obtain estimates of
the lower bounds on the value of $T_c^0$ for copper and gold.
These lower bounds are given in the seventh column of
\tableref{table:experiments}.

These estimates of $T_c^0$ are very sensitive [see
Eqs.~(\ref{eq:lowT2}) and (\ref{conclusions:PCfirst})] to the
experimental parameters. For example, Ref.~\onlinecite{LDDB} points
out that the measured values of the PC are correct up to a factor of
two. The exact minimal value of $T_c^0$ that satisfies
\equref{fittingExperiments:condition} for copper, based on the
values quoted in \tableref{table:experiments}, is $4$mK. However,
assuming half of the value reported in Ref.~\onlinecite{LDDB} for
the PC, results in a minimal $T_c^0$ of about $0.3$mK. The curves in
\figref{fig:Tc0_over_Ec_versus_Ls_over_L} ignore the error bars in
the experiments. Thus, the values of $T_c^0/E_c$ in this figure
should be considered only as rough estimates.

Besides spin-flip scattering from magnetic impurities, decoherence
of the electrons is also caused by other processes, e.g.,
electron-phonon inelastic interactions. Hence $L_s$ is always larger
or of the order of the dephasing length. A lower bound on $L_s$ is
given by equating it to the measured $L_\varphi$. Those values (see
\tableref{table:experiments}) are small enough to fulfil the
condition in \equref{fittingExperiments:condition}. In other words,
we could account for the data of copper and gold since the measured
$L_\varphi$ was small enough. This is not the case for silver,
\cite{DBRBM} where $L/L_\varphi(=0.3)$ is too small to explain the
result $I(T=4.6 E_c)=1.6eE_c$ using \equref{eq:contour}. Our theory
is not applicable to that experiment. We believe that the reason for
that, as explained above, is the high frequency used in that
experiment.

\section{Discussion}
\label{section:results}

In our result for the PC, Eq.~(\ref{eq:contour}), there appears
the bare transition temperature and not the one reduced by the
pair-breaking mechanism. Therefore we propose the scenario, in
which the bulk transition temperature vanishes due to the
pair-breaking mechanism, while the PC is dominated by a
relatively high attractive interaction.

The bulk $T_c$ vanishes due to pair breaking for $L_s<\xi(0)$.
However, we find that the PC may still be hardly affected by
pair breaking. The physical reason for that is that as long as
$L_s > L$ the Cooper pair fluctuations can complete a circle
around the ring before being magnetically scattered, and hence
respond to the Aharonov-Bohm flux. The PC is immune to pair
breaking in the regime given by \equref{2} where the bulk form
is normal. This is demonstrated in Figs.~\ref{fig1} and
\ref{fig2}.

In the pair-breaking regime given by \equref{2}, the upper bound on
the dominant quantum fluctuations ($\nu\neq 0$) is determined by the
Thouless energy. Dominant fluctuations of high Matsubara frequencies
necessitate high wave numbers. Therefore, at low temperature $T\ll
E_c$ high wave numbers are involved too in the dominant fluctuations
[see \figref{fig:qI_of_phi}], in contrast to the effective
dimensional reduction occurring in other phenomena when $L\ll L_T$,
notably weak-localization corrections.\cite{0D} The maximal number
of flux harmonics that contribute to the PC, \equref{max_m}, is
bounded due to thermal fluctuations and due to spin-flip scattering.
Consequently, in a system with magnetic impurities, even at zero
temperature the PC may not have the sawtooth shape, which appears
for the PC's without pair breaking, see \figref{fig:m_harmonics}.

The effective interaction is renormalized upwards with decreasing
energy; For the bulk it stops at $\sim\max\{T_c^0,1/\tau_s\}$ (which
explains why $T_c$ disappears for $1/\tau_s\gtrsim T_c^0$). In the
mesoscopic range, $E_c\gtrsim T_c^0$, the Thouless energy sets
another bound for the energy scale at which the renormalization
stops. In \secref{section:renormalization} it is shown that these
considerations agree with our result for the PC \equref{eq:contour},
see \figref{fig:first}.

We found that in the high-temperature regime, the PC decreases
exponentially with $L/L_s$ or with $L/L_T$, whichever is larger. The
explicit exponential decay of the PC with $L/L_s$ in both the high
and the low temperature regimes [Eqs.~(\ref{eq:very_high_T}) and
(\ref{eq:lowT2}) respectively] for $L\gtrsim L_s$ is in agreement
with the qualitative argument of \equref{2}. Note that
\equref{eq:lowT2} is applicable only at very low temperatures, such
that $T\ll \{T_c^0,E_c\}$. The experiments on copper\cite{LDDB} and
gold\cite{JMKW} rings correspond to $T_c^0\sim 1mK$, thus
\equref{eq:lowT2} can be used only at very low temperatures $T\ll
1mK$. In the experiments the lowest temperature was $\sim 10$~mK,
and therefore the measured PC cannot be precisely fitted by the
approximate expression \equref{eq:lowT2}. In the low-temperature
regime the dependence of the PC on $T_c^0$ is logarithmically weak
[see \equref{eq:lowT2}]. This weak dependence explains why in
Ref.~\onlinecite{AEEPL}, where the transition temperature was taken
as $10\mu$K (in the absence of pair breaking), the result was
smaller only by a factor of $\sim 5$ compared with the
experiment.\cite{LDDB}

Interestingly enough, it follows from our work that by measuring the
PC and the pair-breaking strength one may determine $T_{c}^{0}$,
which would be directly measurable only if enough low-temperature
pair breaking could be eliminated. This elimination is very hard to
achieve in some materials. Our result \equref{MAIN} can explain the
large PC of Refs.~\onlinecite{LDDB} and \onlinecite{JMKW}, with
$L_s$ value larger than (or of the order of) the measured
$L_\varphi$ (see \tableref{table:experiments} and
\figref{fig:Tc0_over_Ec_versus_Ls_over_L}). Even though $L_s$ was
not measured in the PC experiments, we obtain a lower bound on the
bare transition temperatures for copper and gold. These minimal
$T_{c}^0$'s correspond to minimal pair-breaking strength given by
$L_s\sim 5\;\mu$m in the copper sample\cite{LDDB} and $L_s\sim
25\;\mu$m in the gold sample.\cite{JMKW} The fitted maximal $L_s$'s
can be caused by a very low (less than one part per million)
concentration of magnetic impurities. These concentrations seem
appropriate for the purest copper and gold samples available
experimentally.\cite{PGAPEB} Although, a full consideration of the
effect of the magnetic impurities, including Kondo physics, is still
necessary.

Our result concerning the fundamentally different sensitivities
of $T_c$ and PC's to pair breaking is valid regardless of the
situation in specific materials. Our idea can be tested, for
example, by measuring the persistent currents in very small
rings made of a superconducting material whose transition
temperature is known, as functions of possible pair-breaking
mechanisms. For $E_c\gtrsim 100\;$mK, say, and a material with
$T_c^0$ of a few $10\;$mK, the range of pair breaking which
satisfies \equref{1} becomes easier to control experimentally.

\begin{acknowledgments}
We thank E. Altman, L. Bary-Soroker, A. M. Finkel'stein, L.
Gunther, D. Meidan, K. Michaeli, A. C. Mota, F. von Oppen, Y.
Oreg, G. Schwiete, and A. A. Varlamov for very helpful
discussions. This work was supported by the German Federal
Ministry of Education and Research (BMBF) within the framework
of the German-Israeli project cooperation (DIP), by the Israel
Science Foundation (ISF), and by the Emerging Technologies
program.
\end{acknowledgments}

\appendix

\section{Derivation of the partition function}\label{appendix}
Here we derive, using the method of Feynman path integral, the
partition function, \equref{ZwithPi}. In terms of the Grassmann
variables $\psi_\alpha({\bf r},\tau)$ [$\bar\psi_\alpha({\bf
r},\tau)$], the partition function reads
\begin{align}
\mathcal Z=\int D(\psi({\bf r},\tau),\bar\psi({\bf
r},\tau))\exp(-\tilde {\cal S} )\ ,
\end{align}
where the action $\tilde{\cal S}$ is
\begin{align}
\tilde {\cal S}=\int d{\bf r} \int_0^\beta d\tau
\left[\bar\psi_\sigma({\bf r},\tau)\partial_\tau \psi_\sigma({\bf
r},\tau) +\mathcal{H}({\bf r},\tau)\right]\ .
\end{align}
Here $\beta=1/T$ and $\mathcal{H}$ is given by the integrand of
the Hamiltonian, \equref{HAM}, with Grassmann variables (of the
same imaginary time) replacing the creation and annihilation
operators. Introducing the bosonic fields $\Delta ({\bf
r},\tau)$ via the Hubbard-Stratonovich transformation, the
partition function takes the form
\begin{align}\label{appendix:Z}
&\mathcal Z=\int D(\psi({\bf r},\tau),\bar\psi({\bf
r},\tau))D(\Delta({\bf r},\tau),\Delta^{\ast}({\bf r},\tau))
\nonumber\\& \times \exp(-{\cal S})\ ,
\end{align}
where the differential of the bosonic field $\Delta({\bf
r},\tau)$, $D(\Delta({\bf r},\tau),\Delta^{\ast}({\bf r},\tau))$,
contains a factor of $\beta V/\pi g$. The action ${\cal S}$ is
given by
\begin{align}
{\cal S}&=\int d{\bf r}\int_{0}^{\beta} d\tau \Bigl
(\frac{|\Delta ({\bf r},\tau )|^{2}}{g}
\nonumber\\
& -\frac{1}{2}\bar{\Psi}({\bf r},\tau )G^{-1}_{{\bf r},{\bf
r};\tau ,\tau}\Psi({\bf r},\tau )\Bigr )\ ,\label{ACT1}
\end{align}
where
$\bar{\Psi}=(\bar{\psi}_{\up},\bar{\psi}_{\down},\psi_{\up},\psi_{\down})$,
and the inverse Green function $G^{-1}$ (at equal positions ${\bf
r}$ and equal imaginary times $\tau$) is
\begin{widetext}
\begin{align}
G^{-1}_{{\bf r}={\bf r}';\tau =\tau '} =\left
[\begin{array}{cccc}
-\partial_{\tau}-h^{\phi}_{\up}&-2u_{2}S_{-} &0&\Delta\\
-2u_{2}S_{+}& -\partial_{\tau}-h^{\phi}_{\down}&-\Delta &0\\
0&-\Delta^{\ast}&-\partial_{\tau}+h^{-\phi}_{\up}& 2u_{2}S_{+}\\
\Delta^{\ast}&0&2u_{2}S_{-}&-\partial_{\tau}+h^{-\phi}_{\down}\end{array}\right
]\equiv\left [\begin{array}{cccc}\hat{G}^{-1}_{\rm p}&\ &\ &\hat{\Delta}\\
\ &\ &\ &\ \\
\ &\ &\ &\ \\
\hat{\Delta}^{\dagger}&\ &\ &\hat{G}^{-1}_{\rm
h}\end{array}\right ]\ .\label{MAT}
\end{align}
Here $h_{\alpha}^{\pm\phi}={\cal H}_{0}(\pm \phi)+u_{1}+{\rm
sgn}(\alpha )S_{z}u_{2}$, and $S_\pm=(S_x\pm iS_y)/2$, where
${\rm sgn}(\up)=1$ and ${\rm sgn}(\down)=-1$.

The integration over the fermionic part in \equref{appendix:Z}
yields
\begin{align}
{\cal Z}=\int D(\Delta({\bf r},\tau),\Delta^{\ast}({\bf
r},\tau)) \exp \Bigl (\frac{1}{2}{\rm Tr}\ln (\beta G^{-1})-
\int d{\bf r }\int_{0}^{\beta} d\tau\frac{|\Delta({\bf r},\tau
)|^{2}}{g} \Bigr )\ .\label{ZwithTr}
\end{align}
We expand ${\rm Tr}\ln (\beta G^{-1})$ up to second
order\cite{COM3} in $\Delta$
\begin{align}\label{huge}
{\rm Tr}\ln (\beta G^{-1})={\rm Tr}\ln (\beta
G^{-1}_0)-\frac{1}{(\beta V)^2} \iiiint d{\bf r}d{\bf r}'d\tau
d\tau'\; {\rm Tr} \left[\hat G_p({\bf r}',\tau';{\bf
r},\tau)\hat\Delta({\bf r},\tau)\hat G_h({\bf r},\tau;{\bf
r}',\tau')\hat\Delta^\dag({\bf r}',\tau')\right]\ .
\end{align}
\end{widetext}
The inverse Green function for non-interacting electrons,
$G^{-1}_0$, is given by \equref{MAT} for $\Delta=0$. The first
term on the right hand side of Eq. (\ref{huge}), which is
zeroth-order in $\Delta$, gives rise to the partition function of
non-interacting electrons, ${\cal Z}_0=\det(\beta G^{-1}_0)$.

In \equref{huge}, $G_{p}$ ($G_{h}$) is the particle (hole) Green
function. These functions are the solutions of
\begin{align}
\hat G^{-1}_{p/h}({\bf r},\tau) \hat G_{p/h}({\bf r},\tau ;{\bf
r}',\tau')=\delta({\bf r}-{\bf r}')\delta(\tau-\tau')\
,\label{eq:definitionGF}
\end{align}
where $G_{p/h}^{-1}$ are defined in \equref{MAT}. As can be seen
in that equation, the particle and the hole inverse Green
functions are related to one another by
\begin{align}
&\hat G^{-1}_h({\bf r},\tau,\phi,S_+,S_-,S_z)\nonumber\\&= -\hat
G_p^{-1}({\bf r},-\tau,-\phi,S_-,S_+,S_z)\ .\label{eq:inverseGF}
\end{align}
Therefore,
\begin{align}
&\hat G_h({\bf r},\tau;{\bf
r}',\tau',\phi,S_+,S_-,S_z)\nonumber\\&= -\hat G_p({\bf
r},-\tau;{\bf r}',-\tau',-\phi,S_-,S_+,S_z)\nonumber\\&= -\hat
G_p({\bf r},\tau';{\bf r}',\tau,-\phi,S_-,S_+,S_z)\ ,
\label{eq:relationGF1}
\end{align}
where in the last equality we have used time-translational
invariance to shift $\tau$ and $\tau '$ by $\tau +\tau'$.
Reversing the sign of the flux $\phi$ together with
interchanging ${\bf r}$ and ${\bf r}'$ leads to the relation
(the superscript $t$ denotes the transposed matrix)
\begin{align}
&\hat G_p({\bf r},\tau';{\bf
r}',\tau,-\phi,S_-,S_+,S_z)\nonumber\\&=\hat G_p^t({\bf
r}',\tau';{\bf r},\tau,\phi,S_+,S_-,S_z)\ .
\label{eq:relationGF2}
\end{align}
We have used Eqs.~(\ref{eq:relationGF1}) and
(\ref{eq:relationGF2}) to replace the hole Green function in
\equref{huge} by a particle Green function. Then, in momentum
representation, the second term of the right hand side of
\equref{huge} reads\cite{COM2}
\begin{widetext}
\begin{align}
{\rm Tr}\ln (\beta G^{-1})\Big |^{2\textrm{nd}}= \sum_{{\bf
q}_1,{\bf q}_2,\nu}\sum_{{\bf k}_1,{\bf k}_2,\omega} {\rm Tr
}\left[\hat G_{\rm p}({\bf k}_1+{\bf q}_1,{\bf k}_2+{\bf
q}_2,\omega+\nu)\hat\Delta({\bf q}_2,\nu)\hat G_{\rm p}^t(-{\bf
k}_1,-{\bf k}_2,-\omega)\hat\Delta^\dag({\bf q}_1,\nu)\right] \
.\label{SEC}
\end{align}
\end{widetext}
The flux dependence is incorporated into the momenta ${\bf p}$,
where ${\bf p}^2/2m-\mu$ are the eigenvalues of ${\cal
H}_{0}(\phi)$. Thus, the longitudinal components of the momenta
in the Green function $G$ have the form
\begin{align}\label{momentaFermi}
2\pi(n+\phi)/L\ ,
\end{align}
while those of the momenta in the boson field $\Delta $ are
\begin{align}\label{momentaBose}
2\pi(n+2\phi)/L \; ,
\end{align}
where $n$ is an integer. The Matsubara frequencies of the Green
functions, $\omega+\nu$ and $-\omega$, are fermionic $[=\pi(2n+1)
T]$. The order-parameter fluctuations are characterized by the
Matsubara bosonic frequencies $\nu=2\pi n T$.

The resulting expression for the partition function may be
simplified since the terms that survive the disorder-average in
the sum of Eq.~(\ref{SEC}) are those for which \cite{AGD} ${\bf
q}_{1}={\bf q}_{2}$. Following Ref.~\onlinecite{AGD}, we
disorder-average over the exponent in \equref{ZwithTr}, rather
than over the free energy, to obtain an answer which is correct
to leading order in $(\mu\tau_+) ^{-1}$. Finally we trace over
the product of the $2\times 2$ matrices in \equref{SEC} and
integrate over $\Delta$ in \equref{ZwithTr}. In this way we
obtain the partition function, \equref{ZwithPi}.


\begin{thebibliography}{999}
\bibitem{COM3}
This expansion, up to second order in $\Delta$, is valid for
temperatures well above the transition temperature, and,
strictly speaking, above the Ginzburg critical region. In the
case of interest to us, in which $T_c$ vanishes, it is not
obvious that a finite Ginzburg critical region exists at all.
\bibitem{COM2}
The order of the momenta of the second Green function has been interchanged
in Eqs.~(9) and (12) of
Ref.~\onlinecite{BarySoroker_EntinWohlman_Imry}. This typo was
not carried on in the derivation there.
\bibitem{COM4}
We use the fact that any meromorphic function $f$ can be written
locally as $(z-z_0)^n g(z)$, where $g(z_0)\neq 0,\infty$. Hence,
\begin{align}\label{logDerivative}
\frac{f'}{f}=\frac{d}{dz}(\ln f)=\frac{n}{z-z_0}+\frac{g'}{g}\ ,
\end{align}
where $f'/f$ has a simple pole at $z_0$ with a residue $n$.
\bibitem{COM5}
The difference in sign between the zero and the pole terms in
\equref{eq:contour} is due to the different sign of $n$ in Eq.
(\ref{logDerivative}) for the poles and the zeros of the
function in the logarithmic derivative. [We checked numerically
that \equref{MAIN} and \equref{eq:contour} are identical.]
\end{thebibliography}
\end{document}